\documentclass{pazhb_eng}
\usepackage{graphicx}
\usepackage{natbib}
\usepackage{color}

\usepackage{multirow}

\definecolor{darkblue}{rgb}{0,0,0.9}

\def\smfigure#1#2#3{
  \begin{minipage}{1.0\columnwidth}
    \begin{minipage}{0.049\columnwidth}
      \rotatebox{90}{\small\phantom{0000}#3}
    \end{minipage}
    \begin{minipage}{0.95\columnwidth}
      \includegraphics[bb=34 155 578 655,width=0.97\columnwidth]{#1}
      \centerline{\small #2}
    \end{minipage}

    \vskip 3pt
    ~
  \end{minipage}
}

\def\fdg{\hbox{$~\!\!^\circ$}}
\def\à{$^{\mbox{\small a}}$}
\def\á{$^{\mbox{\small b}}$}
\def\â{$^{\mbox{\small c}}$}
\def\ã{$^{\mbox{\small d}}$}

\begin{document}

\journalinfo{2012}{38}{8}{492}[496]

\title{Luminosity Function of High-Mass X-ray Binaries and Anisotropy in the Distribution of Active Galactic Nuclei toward the Large Magellanic Cloud}

\author{A.\,A.\,Lutovinov\email{aal@iki.rssi.ru}\address{1}, S.\,A.\,Grebenev\address{1}, and S.\,S.\,Tsygankov\address{2,3,1}\\
\bigskip
  {\it (1) Space Research Institute, Russian Academy of Sciences, Moscow, Russia}\\
  {\it (2) Finnish Centre for Astronomy with ESO (FINCA), University of Turku, Piikki\"o, Finland}\\
  {\it (3) Astronomy Division, Department of Physics, University of Oulu, Finland}
}

\shortauthor{Lutovinov et al.}

\shorttitle{Luminosity Function of High-Mass X-ray Binaries in LMC}

\submitted{27 March 2012}

\begin{abstract}

In 2003--2012, the INTEGRAL observatory has performed long-term observations of the
Large Magellanic Cloud (LMC). At present, this is one of the deepest hard X-ray (20--60 keV) surveys of
extragalactic fields in which more than 20 sources of different natures have been detected. We present the
results of a statistical analysis of the population of high-mass X-ray binaries in the LMC and active galactic
nuclei (AGNs) observed in its direction. The hard X-ray luminosity function of high-mass X-ray binaries is
shown to be described by a power law with a slope $\alpha\simeq1.8$, that in agreement with the luminosity function
measurements both in the LMC itself, but made in the soft X-ray energy band, and in other galaxies. At the
same time, the number of detected AGNs toward the LMC turns out to be considerably smaller than the
number of AGNs registered in other directions, in particular, toward the source 3C 273. The latter confirms
the previously made assumption that the distribution of matter in the local Universe is nonuniform.

\englishkeywords{hard X-ray sources, high-mass X-ray binaries, active galactic nuclei}

\end{abstract}

\section{Introduction}
\label{sec:intro}

    The all-sky survey that has been performed by
the INTEGRAL observatory since 2003 in the hard
X-ray ($>20$ keV) energy band has allowed one not
only to discover several hundred new X-ray sources
(Krivonos et al. 2007, 2010a; Bird et al. 2010) but also
for the first time to carry out a fairly comprehensive
analysis of the statistical properties of objects of
different classes: active galactic nuclei (Sazonov
et al. 2007, 2008), high-mass and low-mass X-ray
binaries in the inner region of our Galaxy (Lutovinov
et al. 2005; Revnivtsev et al. 2008). Over the
last several years, the INTEGRAL observatory has
performed ultra deep observations of several regions,
virtually reaching the limits of possibilities of
coded-aperture telescopes. The next improvement in the
sensitivity of hard X-ray sky surveys may be expected
only with the advent of grazing-incidence orbital
telescopes with new-generation multilayered mirrors
(NuSTAR, Astro-H, ART-XC/SRG).

   The field toward the Large Magellanic Cloud
(LMC) observed with the INTEGRAL instruments
in 2003--2004 and 2010--2012 for more than 7 Ms (the
effective exposure time was $\sim4.8$ Ms) is among the
regions with deep coverage. The primary target in
these observations was the remnant of Supernova
1987A in order to record the emission lines of
the radioactive decay of $^{44}{\rm Ti}$ synthesized at the time
of its explosion (Grebenev et al. 2012a). However,
such a long exposure time also made it possible to
record more than twenty point hard X-ray sources of
different natures (Grebenev et al. 2012b).

   The goal of this work is a statistical analysis of
high-mass X-ray binaries (HMXBs) in the LMC and
active galactic nuclei (AGNs) registered in its direction
as well as a comparison of our results with those from
observations in other energy bands and in other sky
regions.

   A detailed description of the data from the IBIS
and JEM-X telescopes of the INTEGRAL observatory
(Winkler et al. 2003), a complete list of registered
sources, their identification, etc. are presented in the paper of
Grebenev et al. (2012b). This work is based on the
data from the ISGRI detector of the IBIS telescope
obtained in the 20--60 keV energy band; the image
processing and reconstruction methods are described
in Krivonos et al. (2010b).

\section{Data analysis}
\label{sec:data}

\begin{figure}
\centering

\smfigure{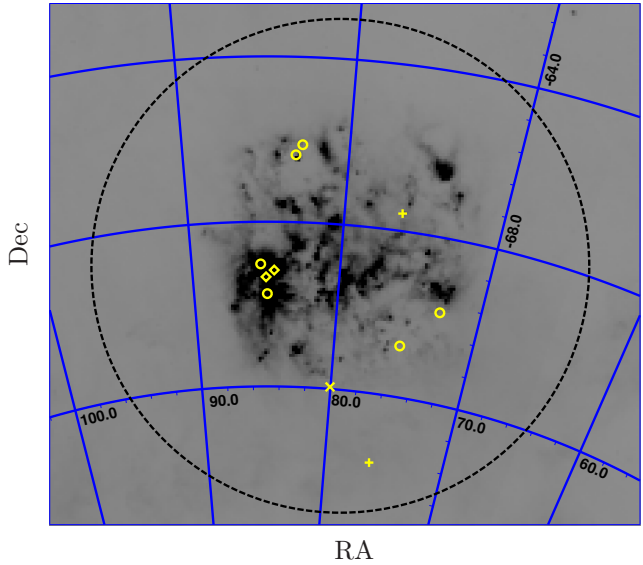}{RA}{Dec}

\caption{\small The $100-\mu$m (IRAS) image of the LMC reflecting
  the dust distribution. The dashed line corresponds to a circle
  with a radius of 6\fdg. The sources detected by the INTEGRAL
  observatory in the 20--60 keV energy band are indicated by different
  symbols: the circles are high-mass X-ray binaries, the crosses are
  active galactic nuclei, the diamonds are single pulsars, and the
  symbol X denotes a low-mass X-ray binary.  }\label{image}

\end{figure}

    The field of view of the IBIS telescope is fairly wide
($29$\fdg$\times29$\fdg). In combination with a special
observing technique, whereby each succeeding pointing is
shifted by $2$\fdg relative to the preceding one (for more
details, see Winkler et al. 2003), it allows large sky
regions to be covered. In particular, to investigate the
population of AGNs, we chose a sky field centered on
the LMC (05$^h$ 24$^m$, -70\fdg) with a radius of 24.5\fdg. In
this sky field, we registered five AGNs with fluxes  $F\gtrsim 6\times10^{-12}$
erg s$^{-1}$ cm$^{-2}$ in the 20--60 keV energy
band on the map averaged over the entire time of our
observations (four of them had a confidence level  $>5\sigma$,
which was the threshold when selecting the sources
for our subsequent analysis)\footnote{It is necessary to note that totally six extragalactic objects were detected in the direction to LMC (Grebenev et al. 2012b), but the flux of one of them is below $6\times10^{-12}$ erg s$^{-1}$ cm$^{-2}$ and the significance of the detection of another one is below $5\sigma$.}. Figure 1 shows an
infrared (IRAS, $100-\mu$m) map of the sky field around
the LMC that reflects the dust distribution in the
galaxy. The positions of the X-ray sources of different
classes registered by the INTEGRAL observatory are
marked in this figure. The LMC proper is located
within the circle with a radius of  6\fdg (dashed line). As
we see from the figure, the INTEGRAL observatory
detected six HMXBs within this region (one more
source is a candidate for objects of this class, but
since there is no ultimate confirmation of its nature, it
was excluded from consideration). All HMXBs were
recorded on the averaged map at a confidence level
$>5\sigma$.

    Before constructing the log(N)-log(S) distribution,
it is necessary to determine the completeness and
sensitivity of the survey. In Fig. 2, the area of
the survey is plotted against the flux from a source
registered at a $5\sigma$ confidence level (a flux of 1 mCrab in
the 20--60 keV energy band corresponds to $\simeq1.2\times10^{-11}$
erg s$^{-1}$ cm$^{-2}$ ) for the above two regions: the
solid line is for the full field 24.5\fdg in radius; the dashed
line is for the LMC itself (within the 6\fdg radius). We see that
the central region has almost uniform coverage down
to fluxes of $\sim0.5$ mCrab. At the same time, in the
wider region, it is important to take into account the
coverage nonuniformity even at fluxes of $\sim 20$ mCrab.
An appropriate correction was done using the method
described by Shtykovskiy and Gilfanov (2005).

\begin{figure}
\centering

\includegraphics[width=\columnwidth,bb=60 275 550 722]{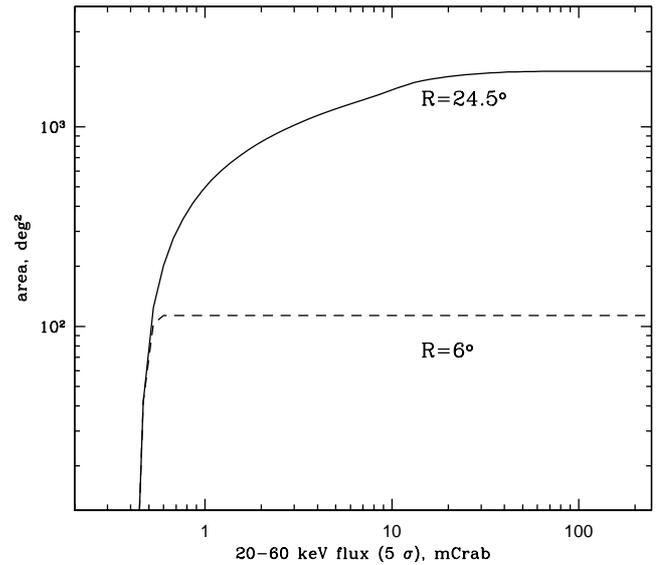}

\caption{\small Area of the survey toward the LMC versus flux
from a source detected at a $5\sigma$ confidence level for regions
with radii of 24.5\fdg (solid line) and 6\fdg (dashed line).
}\label{lmc_cover}

\end{figure}

\section{THE HMXB LUMINOSITY FUNCTION}
\label{sec:lfhmxb}

    As was shown in several papers (see, e.g., Grimm
et al. 2002, 2003; Mineo et al. 2012), the differential luminosity function
of HMXBs in galaxies of different types is proportional
to the star formation rate (SFR) in the galaxy
under consideration and has a universal power-law
form,

$$\frac{dN}{dL}\propto SFR \times L^{-\alpha} \hspace{5cm} (1)$$

with a slope $\alpha\simeq1.6\pm0.1$ in the wide luminosity
range $10^{34}-10^{40}$ erg s$^{-1}$. This is most likely
explained by the fundamental mass-luminosity and
mass-radius relations for high-mass stars (Postnov
2003). Thus, the luminosity function is an
important tool in investigating both the formation
and evolution of binary systems and the influence of
different metallicities and chemical compositions in
different galaxies on this process (Shtykovskiy and
Gilfanov 2005). On the other hand, the luminosity
function has usually been and is being constructed
in the soft X-ray 1--10 keV energy band, where
modern telescopes are sensitive enough to detect
a considerable number of objects in neighboring
galaxies. However, the luminosity in this energy
band constitutes a small fraction of the total
luminosity of usually hard and absorbed HMXBs, which can introduce
distortions into the derived relations. Therefore,
investigating the hard X-ray luminosity function of
binaries and comparing it with the results obtained
previously are of significant interest, despite the
much smaller number of recorded sources and, as a
result, the poorer statistics.

    Using the maximum likelihood estimation, we can
fit the differential luminosity function of HMXBs in
the LMC by the power law (1) with a slope $\alpha=1.8^{+0.4}_{-0.3}$
and an expected number of high-luminosity
sources $N (L_X\gtrsim2\times10^{36}$ erg s$^{-1}$) $\approx9$.
The corresponding cumulative luminosity function is indicated
by the histogram in Fig. 3. For comparison, the
same figure shows the luminosity functions with
$\alpha=1.28^{+0.26}_{-0.23}$ and $N (L_X\gtrsim10^{35}$ erg s$^{-1}$) $\approx5$ measured
from XMM-Newton data (gray lines, Shtykovskiy and Gilfanov 2005) and predicted
from the "universal" luminosity function with
$\alpha\approx1.6$ and $N (L_X\gtrsim10^{35}$ erg s$^{-1}$) $\approx11\pm5$ (black lines,
Grimm et al. 2003).
It should be noted that both these functions and
their normalizations were obtained in the soft X-ray
 2--10 keV energy band for the LMC region observed
by the XMM-Newton observatory. To compare them
with the luminosity function obtained by
the INTEGRAL observatory in the hard 20--60 keV
energy band for the entire LMC, we made appropriate
renormalizations by assuming the flux ratio for the
spectra of typical HMXBs with neutron stars (which
are the majority among the objects of this class;
Lutovinov et al. 2007) to be $F_{2-10 {\rm keV}}/F_{20-60 {\rm keV}}\simeq0.5$
(see, e.g., Filippova et al. 2005) and the corresponding
star formation rates in the entire LMC and
the region covered by the XMM-Newton observatory
to be $SFR(LMC)\approx0.5$ $M_{\odot}$ yr$^{-1}$ and
$SFR({\rm XMM})\approx0.089$ $M_{\odot}$ yr$^{-1}$ , respectively (Shtykovskiy and
Gilfanov, 2005). We see from the figure that the
INTEGRAL measurements in the range of high
luminosities agree satisfactorily both with the soft
X-ray results for LMC and with the predictions following from
the universality of the HMXB luminosity function.
Nevertheless, it should be noted that objects with
"atypical" spectra (including black holes and neutron
stars as compact objects) are encountered among
HMXBs, which can lead to some distortions of the
luminosity function when passing from one energy
band to the other.

\begin{figure}
\centering

\includegraphics[width=\columnwidth,bb=55 240 550 700]{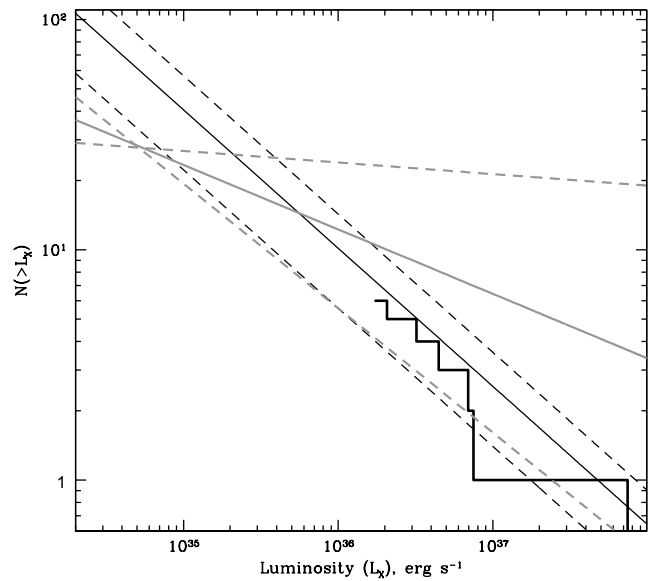}

\caption{\small The cumulative luminosity function of HMXBs in
the LMC from INTEGRAL data (histogram). The gray
and black solid lines indicate the luminosity functions
measured from XMM-Newton data for the LMC and a
large sample of galaxies, respectively (for more details,
see the text). The dashed lines show the regions of the
corresponding uncertainties.
 }\label{lf_hmxb}

\end{figure}

\section{Log(N)-Log(S) distribution of AGN}
\label{sec:lfagn}

    As has been noted above, ultradeep surveys of
various sky fields, which allow the detection limit
of point sources to be reached with coded-aperture
telescopes, are also used to study the distributions of
objects of different classes over the sky. In particular,
Krivonos et al. (2007) showed that the distribution of
AGNs over the sky is not isotropic -- the number of
AGNs toward the Virgo cluster of galaxies and the
Great Attractor exceeds significantly their number
in the opposite direction. However, the amount of
data available at that time allowed this result to be
obtained on a spatial scale of tens of thousands of
square degrees. Several series of deep observations of
the LMC and the sky field around the source 3C 273
performed in 2008--2012 allowed us to narrow the
characteristic scale of the sought-for asymmetry to
several hundred or thousand square degrees.

\begin{figure}
\centering

\includegraphics[width=\columnwidth,bb=60 275 550 727]{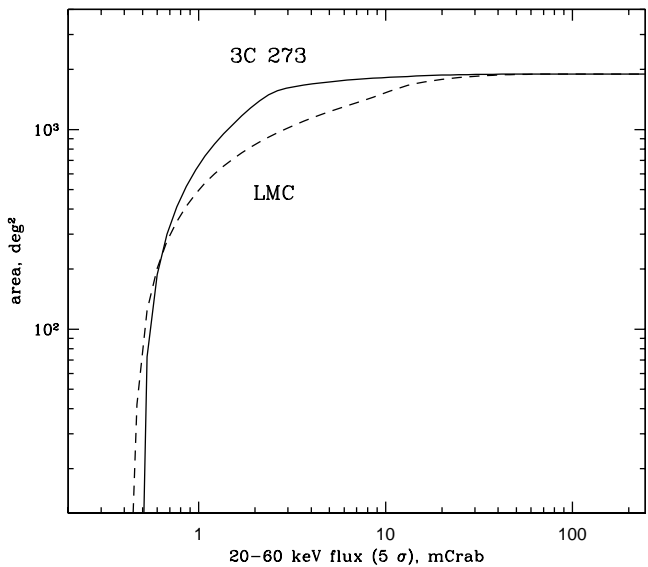}

\caption{\small Comparison of the survey areas around the source
3C 273 (solid line) and the LMC (dashed line) as functions
of the flux from a source recorded at a $5\sigma$ confidence
level.}\label{agns_cover}

\end{figure}
\begin{figure}
\centering

\includegraphics[width=\columnwidth,bb=50 250 550 710]{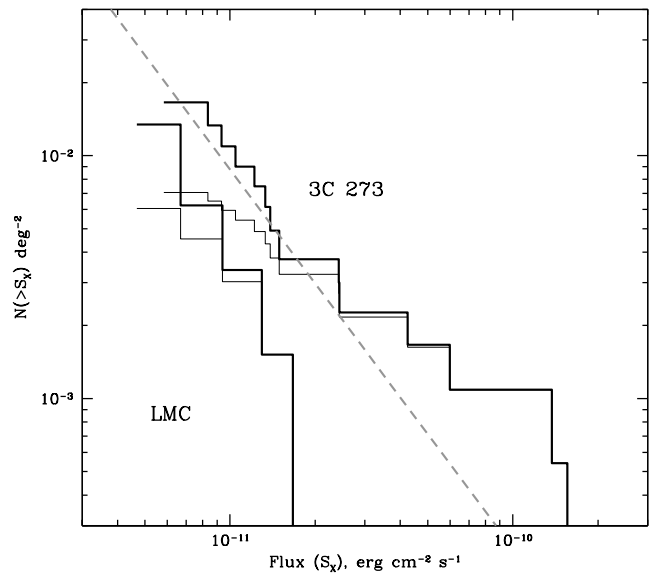}

\caption{\small Cumulative log(N)-log(S) distribution of AGN in ultradeep fields
toward the LMC and 3C 273 (thick histograms). The
thin histograms indicate the observed distribution uncorrected
for the corresponding areas of the survey. The dashed
line indicates the best fit of the number of AGNs in the
entire sky in the 17--60 keV energy band from Krivonos et
al. (2010a) renormalized to the 20--60 keV energy band.}\label{lognlogs}

\end{figure}
    Figure 4 shows the areas of the surveys for the
LMC and the sky field around 3C 273 as functions of
the flux from a source detected at a $5\sigma$ confidence level
(similar to Fig. 2). We see that the areas of the surveys
coincide at high fluxes; however, differences associated
with different strategies of the observations begin
already at fluxes of $\sim30$ mCrab.

   The log(N)-log(S) distributions of AGN in both deep fields are
shown in Fig. 5. It is seen that in the sky field toward
3C 273, the number of AGNs detected at a statistically
significant level ($>5\sigma$) (13 objects) with fluxes
$F\gtrsim 6\times10^{-12}$ erg s$^{-1}$ cm$^{-2}$ is more than a factor
of 3 larger than that toward the LMC (4 objects).
The approximation of the log(N)-log(S) distribution of AGN toward 3C 273 by
the function $N(>S)=AS^{-\beta}$  using the maximum
likelihood method gives a slope $\beta=1.65^{+0.31}_{-0.28}$ and an
expected number of sources $N\approx10$ with fluxes above
1 mCrab in the 20--60 keV energy band within the
circle 24.5\fdg in radius. A similar analysis for the sky
field toward the LMC gives an expected number of
sources $N\approx1$ with a flux above 1 mCrab in the 20--60
 keV energy band within the circle 24.5\fdg in radius.
For comparison, it should be noted that under the
assumption of a uniform distribution of AGNs over
the sky, $N\approx6$ sources with fluxes above 1 mCrab are
expected in the circle with a radius of 24.5\fdg in the 20--60
 keV energy band (Krivonos et al. 2010a).

Note that at low fluxes ($F<2\times10^{-11}$ erg s$^{-1}$ cm$^{-2}$),
the log(N)-log(S) distribution in
the region of 3C 273 is in good agreement with the
same distribution averaged over the entire sky. At the
same time, this field exhibits a clear excess of bright
objects, with most of them having low redshifts and
being closer than $\approx70$ Mpc. The latter confirms the
previously made assumption that the local Universe
is inhomogeneous.

\section{Conclusions}
\label{sec:concl}

   Here, we performed a statistical analysis of the
population of HMXBs in the LMC and AGNs registered
in its direction. We showed that the hard X-ray (20--60
keV) luminosity function of HMXBs could be
fitted by a power law with a slope $\alpha=1.8^{+0.4}_{-0.3}$.
This result is
in agreement both with the measurements of the
luminosity function for HMXBs in the LMC itself, but
in the soft X-ray (1--10 keV) energy band, and with
the predictions derived from a detailed analysis of a
large number of different galaxies.

    At the same time, the number of detected AGNs
toward the LMC turns out to be considerably smaller
than the number of such objects registered in other
directions, in particular, toward the source 3C 273.
This, along with the excess of bright objects at low
redshifts in this direction, confirms the previously
made assumption that the mass in the local Universe
is distributed nonuniformly (see, e.g., Krivonos
et al. 2007).

\bigskip
~\bigskip

\acknowledgements

   We are grateful to M.G. Revnivtsev and R.A. Krivonos
 for helpful discussions. This work was financially
supported by the Russian Foundation for
Basic Research (project nos. 10-02-1466, 11-02-01328,
11-02-12285-ofi-m-2011, 12-02-01265),
the "Nonstationary Phenomena in Objects of the
Universe" Program, the Program for Support of
Leading Scientific Schools of the President of the
Russian Federation (NSh-5603.2012.2), and the
State contract no. 14.740.11.0611.

\parindent=0mm

1. A. Bird, A. Bazzano, L. Bazzani, et al., Astrophys. J. Suppl. Ser. 186, 1 (2010).

2. E. V. Filippova, S. S. Tsygankov, A. A. Lutovinov, and R. A. Syunyaev, Astron. Lett. 31, 729 (2005).

3. S. Grebenev, A. Lutovinov, S. Tsygankov, and C. Winkler, (2012a), in press.

4. S. Grebenev, A. Lutovinov, S. Tsygankov, and I. Mereminskiy, Mon. Not. R. Astron. Soc (2012b), submitted, [arXiv:1207.1750]

5. H.-J. Grimm, M. Gilfanov, and R. Sunyaev, Astron. Astrophys. 391, 923 (2002).

6. H.-J. Grimm, M. Gilfanov, and R. Sunyaev, Mon.
   Not. R. Astron. Soc. 339, 793 (2003).

7. R. Krivonos, M. Revnivtsev, A. Lutovinov, et al., Astron. Astrophys. 475, 775 (2007).

8. R. Krivonos, S. Tsygankov, M. Revnivtsev, et al.,
   Astron. Astrophys. 523, A61 (2010a).

9. R. Krivonos, M. Revnivtsev, S. Tsygankov, et al.,
   Astron. Astrophys. 519, A107 (2010b).

10. A. Lutovinov, M. Revnivtsev, M. Gilfanov, et al., Astron. Astrophys. 444, 821 (2005).

11. A. Lutovinov, M. Revnivtsev, M. Gilfanov, and
    R. Sunyaev, in The Obscured Universe, Proceeding
    the VI INTEGRAL Workshop, Ed. by S. Grebenev,
    R. Sunyaev, and C. Winkler, ESA SP-622 (2007), p.
    241.

12. Mineo S., Gilfanov M., Sunyaev R. MNRAS, 419, 2095 (2012).

13. K. A. Postnov, Astron. Lett. 29, 372 (2003).

14. M. Revnivtsev, A. Lutovinov, E. Churazov, et al.,
    Astron. Astrophys. 491, 209 (2008).

15. S. Sazonov, M. Revnivtsev, R. Krivonos, et al., Astron. Astrophys. 462, 57 (2007).

16. Sazonov S., Krivonos R., Revnivtsev M., et al., Astron. Astrophys. 482, 517 (2008)

17. P. Shtykovskiy and M. Gilfanov, Astron. Astrophys.
    431, 597 (2005).

18. C. Winkler, T. Courvoisier, G. Di Cocco, et al., Astron. Astrophys. 411, L1 (2003).

\end{document}